\documentclass{article}

\usepackage[ruled,linesnumbered]{algorithm2e}

\usepackage{threeparttable}
\usepackage{amsmath,amssymb}
\usepackage{graphicx}
\usepackage{epstopdf}
\usepackage{PRIMEarxiv}
\usepackage{enumitem}
\usepackage[utf8]{inputenc} 
\usepackage[T1]{fontenc}    
\usepackage{hyperref}       
\usepackage{url}            
\usepackage{booktabs}       
\usepackage{amsfonts}       
\usepackage{nicefrac}       
\usepackage{microtype}      
\usepackage{lipsum}
\usepackage{fancyhdr}       
\usepackage{graphicx}       
\graphicspath{{media/}}     
\usepackage{cite}

\title{D-HAL: Distributed Hierarchical Adversarial Learning for Multi-Agent Interaction in Autonomous Intersection Management
}

\author{
  Guanzhou Li \\
  Tsinghua University \\
  Beijing, China\\
  \texttt{ligz19@mails.tsinghua.edu.cn} \\
  \And
  Jianping Wu* \\
  Tsinghua University \\
  Beijing, China\\
  \texttt{jianpingwu@tsinghua.edu.cn} \\
  \And
  Yujing He \\
  Tsinghua University \\
  Beijing, China\\
  \texttt{hyj19@mails.tsinghua.edu.cn}\\
}

\begin{document}
\maketitle

\begin{abstract}
Autonomous Intersection Management (AIM) provides a signal-free intersection scheduling paradigm for Connected Autonomous Vehicles (CAVs). Distributed learning method has emerged as an attractive branch of AIM research. Compared with centralized AIM, distributed AIM can be deployed to CAVs at a lower cost, and compared with rule-based and optimization-based method, learning-based method can treat various complicated real-time intersection scenarios more flexibly. Deep reinforcement learning (DRL) is the mainstream approach in distributed learning to address AIM problems. However, the large-scale simultaneous interactive decision of multiple agents and the rapid changes of environment caused by interactions pose challenges for DRL, making its reward curve oscillating and hard to converge, and ultimately leading to a compromise in safety and computing efficiency. For this, we propose a non-RL learning framework, called Distributed Hierarchical Adversarial Learning (D-HAL). The framework includes an actor network that generates the actions of each CAV at each step. The immediate discriminator evaluates the interaction performance of the actor network at the current step, while the final discriminator makes the final evaluation of the overall trajectory from a series of interactions. In this framework, the long-term outcome of the behavior no longer motivates the actor network in terms of discounted rewards, but rather through a designed adversarial loss function with discriminative labels. The proposed model is evaluated at a four-way-six-lane intersection, and outperforms several state-of-the-art methods on ensuring safety and reducing travel time.
\end{abstract}

\keywords{Hierarchical Adversarial Learning, D-HAL, Autonomous Intersection Management (AIM), learning-based AIM, unsignalized intersection, Connected Autonomous Vehicle (CAV) }

\section{Introduction}
Intersections have the characteristic of intensive interaction, where traffic flows from multiple directions intertwine and multiple types of traffic participants co-exist. The more conflict points than other road structures make intersection an accident-prone scenario. Statistics from U.S. Federal Highway Administration exhibit forty percent of crashes were related to intersections. In addition, intersections are critical nodes in traffic management since waiting at signals and intersection congestion account for a significant portion of travel delays, and the congestion that begins at an intersection can spill over to downstream roadways or even entire road network. Despite decades of development in the era of intelligent transportation systems, the efficiency of existing intersection management techniques (i.e, traffic signals) is still compromised by the stop-and-go waves of traffic flows in front of the stop line, and the ideal green wave pass is hard to fully realize in practice. With the advent of connected autonomous vehicles (CAVs), autonomous intersection management (AIM) demonstrates a paradigm for efficient communication-based self-organizing scheduling at intersection. Unlike human-driven vehicles (HVs) that stop until phase turns green, CAVs achieve orderly passage through intersections by communicating with the intersection manager (IM) and with each other. The communication, namely Vehicle-to-Everything (V2X), including Vehicle-to-Vehicle (V2V) and Vehicle-to-Infrastructure (V2I), allows IM or CAV to get global information about intersections earlier and make far-sighted decisions.

According to the decision party of right-of-way, AIM can operate scheduling policy in a centralized or distributed manner. In the centralized AIM, the intersection manager (IM) takes charge of priority assignment and crossing order optimization; Instead, in the distributed AIM, multiple individual controllers (i.e., CAVs), namely agents, negotiate with each other through V2V to determine the optimal passing plans for themselves. In some cases, roadside units (RSUs) also assist CAV to collect and broadcast messages in the distributed AIM, but the final decision-makers are these agents. For large cities, the centralized AIM requires redeployment of massive computing power to meet the scheduling needs of each intersection, while the distributed AIM can take full advantage of the CAVs' own intelligent system. Furthermore, global control in centralized AIM requires more stringent details of each vehicle's movement and fulfill of more accurate communication and clock synchronization. Hence, we adopt the distributed AIM in this study.

In the framework of centralized and distributed AIM, the scheduling techniques can be categorized into rule-based, optimization-based, and learning-based. The prior rules in rule-based AIM do not guarantee an optimal solution, and the computation complexity of optimization-based method increase significantly as the number of vehicles grows and might not be solvable within linear time. Learning-based method have achieved success in many areas of autonomous driving, and also provide a flexible and easy-to-deploy framework for AIM. Most existing learning-based methods adopt multi-agent reinforcement learning (MARL) to solve the problem of CAV interaction at intersection, and MARL can be modeled as homogeneous or heterogeneous agent problem, where homogeneous agents employ a consistent reasoning model to generate their actions and heterogeneous agents vice versa. In AIM, heterogeneous agent problem faces the challenges of dynamic number of agents and higher interaction complexity, while in the homogeneous agent model, the actions selected from the identical strategy are prone to dilemma decision scenarios, and the simultaneous movements of all CAVs in the absence of information about other agents further lead to conflicts and even collisions.

Treating the homogeneous agent problem as a self-playing process, the same reasoning model actually means that each agent "potentially" knows how the other agents behave, while MARL involves both temporal evolution of the environment and interaction processes, and the large state-action space introduces huge uncertainty, which weaken agent's "potential self-awareness" ability and makes the learning hard to converge. In fact, during the multi-agent interaction at the intersection, we mainly focus on the outcome of each-step interaction and whether the final trajectory conflict. Hence instead of discounting the future reward to current moment using Bellman Equation, we adopt hierarchical adversarial learning to guide the long-term and short-term behaviors of CAVs. The hierarchical adversarial learning involves immediate interaction discriminator and final trajectory discriminator, where the former arbitrates the priority of each-step interaction and the latter qualifies the entire trajectory. The main contributions of this paper includes:

\begin{enumerate}
    \item A hierarchical adversarial learning by long-term and short-term discrimination is adopted to realize the complicated interaction among CAVs, which does not depend on design of reward and has low training cost.
    \item Potential energy method is applied to balance speed acquisition (efficiency) and conflict-avoidance (safety) in different distances away from the stop-lines.
    \item the proposed model is evaluated in a four-way-six-lane intersection and found able to prevent collision and eliminate dead-lock at the crossing zone, and it performs better in safety, efficiency, fairness than the baselines.
\end{enumerate}

The rest of the paper is organized as follows. Section II reviews relevant literature in AIM problem, Section III illustrates details of D-SAL. In Section IV, we conduct experiments to evaluate the performance of D-SAL and compare it with baselines. Conclusions and discussions are presented in Section V.

\section{Literature Review}
The framework of Autonomous Intersection Management (AIM) presented by Drenser et al. laid the cornerstone for investigating the interaction of CAVs at intersections \cite{dresner2004multiagent}. Depending on the decision variables, the AIM can be categorized into priority-based and action-based management. As the name implies, priority-based management produces a crossing order for all approaching vehicles and determines the priority between pairs of conflicting vehicles, while action-based management generates reasonable speed trajectories for CAVs to achieve Pareto optimality among goals like safety, efficiency and fairness. Compared with the action-based approach, the priority-based approach reduces the solution space of the AIM problem, and it can prevent collisions and deadlocks effectively under the restriction of yield-and-go rules. However, action-based approach is more executable in distributed AIM, and the more precise control objectives in action-based way enable more optimal scheduling. In order to combine the advantages of both priority-based and action-based approaches, the hierarchical model employs an upper layer to negotiate priorities and a lower layer to generate optimal velocity trajectories\cite{wang2021game}. The model proposed in our study is also a combination of priority-based and action-based approaches, where the final decision generated by the model is the action of each CAV, but the priority between conflicting vehicles is taken into account during the learning process.

In both priority-based and action-based AIM, each vehicle approaching the intersection acquires a spatiotemporal slot for passage, called the "reservation" for specified CAV. The CAv can safely cross the intersection only when the gap between reservations of other vehicles can accommodate the reservations of ego vehicle without overlap. The success of reservation is affected not only by the arrival time of vehicles in the conflicting direction, but also by the movement status of the front vehicle. In other words, the time heading with the front vehicle restrict the action range of follower. Therefore, platoon-based AIM groups vehicles in the same lane into a queue with constant headway\cite{cao2021platoon}, where the leader in platoon negotiates with other vehicles and intersection manager (IM) for scheduling, and the inter-platoon coordination maintain the stability of platoon. Therefore, platoon-based control can significantly reduce the communication load and fuel consumption. As a fundamental variable in the platoon-based control, the platoon size will affect the stability of platoon and the efficiency of passage in AIM, which is systematically analysed in \cite{zhou2021analytical}. And Li et al. adopted Deep Q learning to determine the optimal platoon size in AIM \cite{li2022modeling}. Apart from the platoon formation among the adjacent vehicles in the same lanes, virtual platoon maps two-dimensional traffic flows in AIM to one-dimensional queue according to their distance to the center of the intersection, and groups non-conflicting vehicles to cross the intersection simultaneously\cite{qian2014priority, zhou2022cooperative}. As opposite to the platoon-based management, individual navigation controls the action of each CAV independently. It has a higher degree of freedom than platoon-based way. The higher solution space poses a challenge for solving the intersection scheduling problem, but it also allows for more detailed control to achieve better performance. Platoon-based control can be considered a special case of independent navigation, where vehicles have the flexibility to weave through intersection at low traffic volumes and form platoon of a certain size at high traffic volumes, and this phenomenon is found in our experiments.

Specifically, the scheduling strategy in AIM reschedule the order of vehicles to pass through intersection, which can be divided into: rule-based approach, optimization-based approach, and learning-based approach. Rule-based scheduling determines the order of vehicles through prior rules and has the advantage of low computational effort. First-Come-First-Serve (FCFS) is one of the earliest and most widely-used rule based scheduling \cite{dresner2008multiagent}. It is convenient to implement and shows enough fairness. However, in some cases like high traffic volumes, FCFS might cause longer passage time than signalized control\cite{levin2016paradoxes}. The Longest-Queue-First (LQF) always allows the vehicles in the direction with longest queue to cross the intersection first, which originated as a pressure-based signal control \cite{wunderlich2008novel} and was transferred to AIM with V2I communication\cite{ghaffarian2012vehicular}. Besides, some researches shows that the rotating through intersections enables near-optimal scheduling\cite{lu2022autonomous}. Rhythm controls sets rhythms for vehicular movements in each direction to stagger through the intersection seamlessly\cite{chen2021rhythmic}. The rule-based approach performs well in simple scenarios, but the hard-coded rules not necessarily lead to optimal solutions in some complex scenarios\cite{lu2022autonomous}.

Another method for AIM problem, optimization-based method, can be classified into four categories: dynamic programming, tree search and graph representation, market-based scheduling, and game theory. Dynamic programming formulates AIM as a mixed integer linear programming (MILP) problem containing an objective function and constraints, where the objective function includes passage efficiency, comfort, and energy consumption, and the constraints describe the physical restriction like non-collision and vehicle dynamics. Many researches adopted techniques like big-M method\cite{jin2012multi}, Brand-and-Bound\cite{yang2016isolated}, and inexact Newton method \cite{jiang2017distributed} to find the optimal solution, and some interesting algorithms are introduced to address the scheduling problem, like machine-job distributor\cite{wu2009intersection}, traveling salesman problem\cite{wu2012cooperative}. Nevertheless, in AIM, the behavior of each vehicle is an independent optimizable variable, thus dynamic programming confronts the challenge of combinational explosion, and experiments exhibited that the MILP can only optimize the crossing order of up to $30$ vehicles in real time\cite{levin2017conflict}. 

Tree search gives a solution to find a near-optimal schedule in feasible time, where the leaf nodes of tree enumerates all the possible schedules and the invalid nodes will be removed based on certain rules. It was first presented in \cite{li2006cooperative}, and subsequent studies tried more efficient tree construction and search including red-black tree\cite{hubmann2017decision}, adaptive belief tree\cite{choi2019reservation}, and Monte-Carlo tree search (MCTS) \cite{xu2019cooperative}. The graph-based AIM models the problem of passing order as a graph, where the nodes represent vehicles or states of vehicles and the edges represent the precedence\cite{ahmane2013modeling,wu2014cooperative,liu2017distributed,lin2019graph,chen2021conflict}. The core problem of this method is to decide whether there are edges between nodes and the direction of the edges (i.e., the "leader-follower relationship), and to cluster nodes to layers where vehicles can cross the intersection simultaneously\cite{liu2017distributed,chen2021conflict}. The graph-based method was exhibited feasible in efficiency improvement and deadlock elimination\cite{lin2019graph}.

The market-based method introduces the concept of "time value" in the AIM problem and differentiates individual willingness to pay for delays. The co-utility maximization asks CAVs to report their utility to intersection manager when they approach the intersection and assigns priorities to maximize the collective welfare, and pair-wise swap occurs between adjacent vehicles in the passage sequence when collective welfare increases\cite{buckman2019sharing,sayin2018information}. Besides, Social Value Orientation (SVO) was adopted to represent the additional delay that can be tolerated by individuals, helping to reduce the gap between the individual goals of the quickest passage and the global goals of congestion reduction\cite{buckman2019sharing}. The other branch of market-based method is auction system, where vehicle obtains intersection reservations by outbidding its rivals\cite{vasirani2012market}. The system budget was adopted to pay for the low-budget vehicles to avoid infinite waiting and ensure fairness\cite{carlino2013auction}. Second-price sealed auction in Vickrey-Clarke-Grove was leveraged to reflect the true value of the spatio-temporal resources at intersections\cite{sayin2018information}. In addition, WIN-FIT offered a collective bidding solution to the problem of rear vehicle in limited lane width being limited by the bids of the front vehicle\cite{chen2015win}.

The game-theoretic system consists of decision makers, a set of candidate actions, and a reward function, where decision makers quantify the value of potential interactions\cite{rahmati2021helping}. The games in AIM can be divided into cooperative games and unilateral games\cite{bressan2010noncooperative}, the cooperation in the games can be reached by sharing benefits needless of external force\cite{stryszowski2020framework}, while information sharing and mutual trust help to achieve the equilibrium of the game\cite{daskalakis2009complexity}. Apart from the game of priority in the conflict zone, the game-in-game presented a double layer model involving the game of platoon formation \cite{wei2018intersection}. Existing game theory tends to build theories for two-player games, and multi-player games are usually decomposed into several two-player games, where the worst outcome is selected as the payoff of one round, the effectiveness and convergence of which are still debatable in large scale multi-party games.

The optimization-based method inevitably has idealized assumptions in the modeling process, and deviations in the actual execution of the CAVs can lead to the need for rescheduling. In contrast, the learning-based method is more flexible in execution. The dominant method in learning-based AIM is multi-agent reinforcement learning (MARL), including Deep Q Learning (DQN)\cite{isele2018navigating,mokhtari2021safe,wang2021reinforcement,karthikeyan2022autonomous,gunarathna2022intelligent}, Dueling Deep Q Learning (DDQN)\cite{shu2021driving}, Deep Deterministic Policy Gradient (DDPG)\cite{li2022continuous}, Proximal Policy Optimization (PPO)\cite{quang2020proximal,shi2022control,xue2022multi}, Twin Delayed Deep Deterministic Policy Gradient (TD3)\cite{liu2021reinforcement,bautista2022autonomous,guillen2022learning}, and Soft Actor-Critic (SAC)\cite{ren2020improving}. Besides, DCL-AIM introduced coordinate state and independent state for CAVs to react in different scenarios\cite{wu2019dcl}, RAIM \cite{guillen2020raim} and adv.RAIM \cite{guillen2022multi} applied encoder-decoder structure with LSTM cell, AIM5LA further considered communication delay based on the adv.RAIM\cite{antonio2022aim5la}, and game theory was utilized to determine the leader-follower to enhance the performance of reinforcement learning in \cite{li2020game,ren2020improving}. Besides, attention mechanism\cite{xue2022multi} and graph convolution network\cite{chen2021graph} were also combined with reinforcement learning in AIM. We proposed a novel learning-based method other than that reinforcement learning in AIM, the reasons and advantages will be given in the next Chapter.

\section{Methodology}
\subsection{Problem Statement}
\begin{figure}[h]
    \centering
    \includegraphics[width=0.6\textwidth]{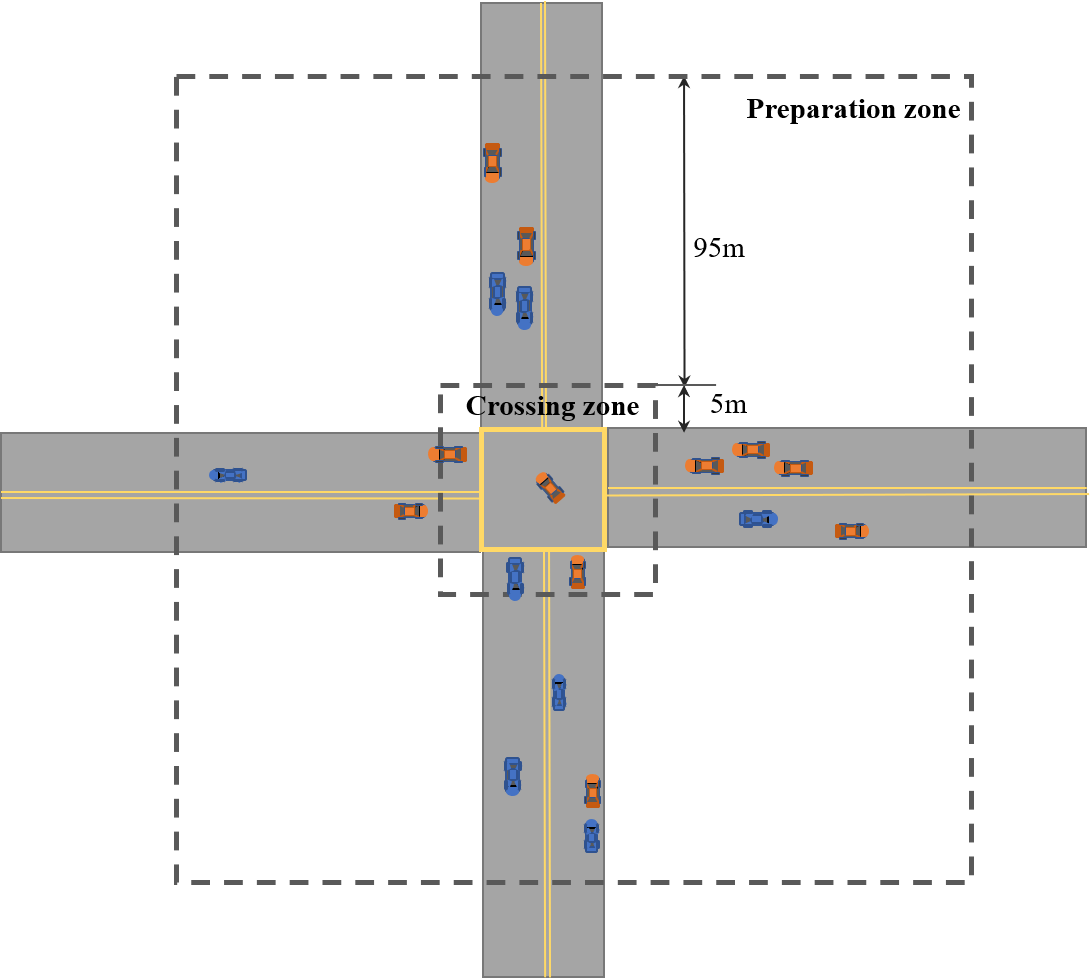}
    \caption{Intersection diagram}
    \label{fig:fig1}
\end{figure}

In this study, we explore a distributed learning-based scheduling at a unsignalized intersection with complete CAVs. The intersection area is partitioned into two zones – preparation zone and crossing zone, which are squares with boundaries 100m and 5m away from the stop lines, respectively, as illustrated in figure \ref{fig:fig1}. And it should be highlighted that the crossing zone is not involved in the preparation zone. In the preparation zone, the navigation algorithm controls each CAV’s next-step action based on its current state so that it can reach the crossing zone at the right time. Before entering the crossing zone, the algorithm will give the suggested crossing speed, and the CAV will pass the crossing zone at this constant speed provided that no conflicts will occur, otherwise it will wait in the preparation zone for the next suggested speed until it can cross safely. After entering the crossing zone, CAVs will not change their speed because frequent speed changes in this zone not only tend to cause collisions, but also have a limited improvement on the passage efficiency.

There are two core problems to be addressed in this study. The first problem is how to adjust the real-time speed for the CAV in the preparation zone to be staggered in arrival time with the incoming traffic in the conflict direction. Besides, under the premise of safety, the scheduling solution should allow CAVs to cross the intersection as rapidly as possible to mitigate delays. The second problem is to provide CAVs with the speeds at which they can pass the crossing zone safely and prevent queuing of vehicles outside the boundary of the crossing zone. It is worth mentioning that this study uses the same actor network to solve both problems, which continuously provides guidance speed to CAVs in the preparation zone.

\subsection{Definitions and Conceptual Designs}
\begin{figure}[h]
    \centering
    \includegraphics[width=\textwidth]{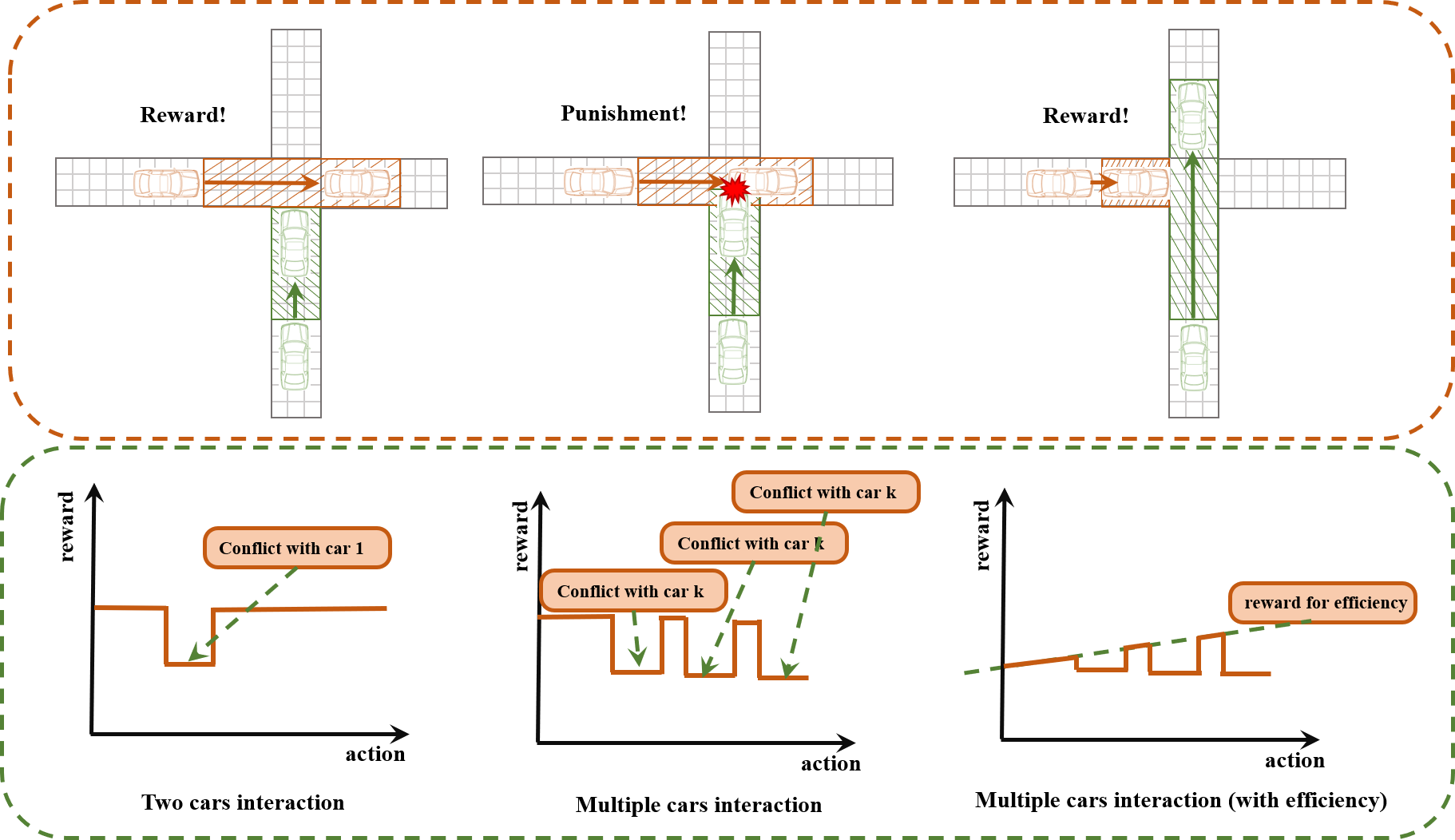}
    \caption{The relationship between actions and rewards}
    \label{fig:fig2}
\end{figure}

\begin{figure}[h]
    \centering
    \includegraphics[width=0.6\textwidth]{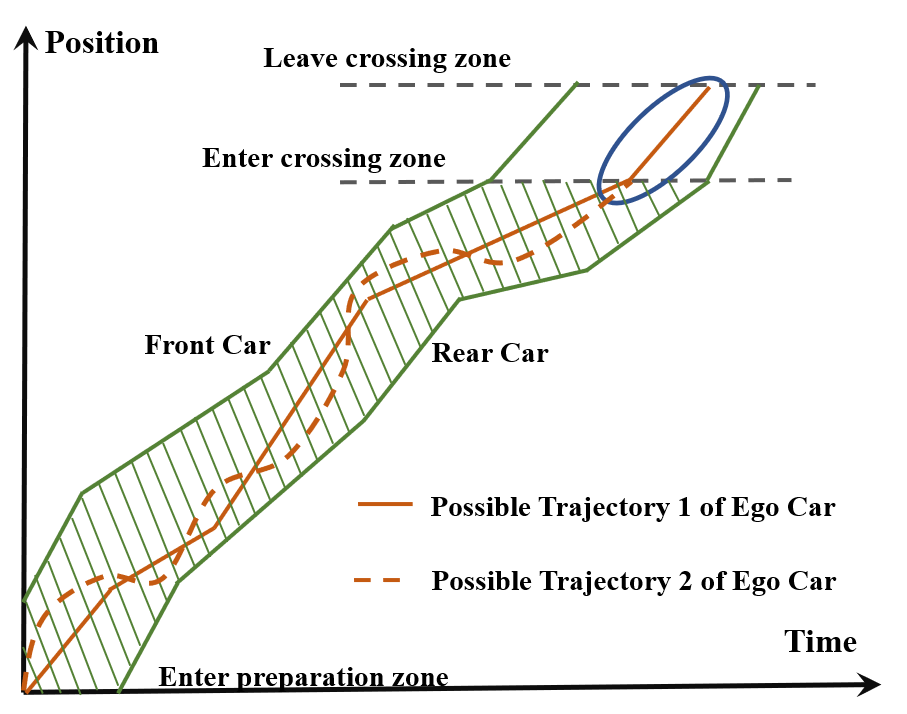}
    \caption{Potential speed trajectories of ego CAVs}
    \label{fig:fig3}
\end{figure}

As discussed in the previous chapter, this study intends to apply a learning-based method to solve the scheduling problem in AIM. In the reinforcement learning framework, CAVs will be rewarded when they passing the intersection safely, and punished otherwise, as shown in figure \ref{fig:fig2}. Under this reward mechanism, each potential interaction forms a notch in the action-reward distribution of the ego CAV, and the interactions with multiple CAVs at different conflict points further form a dentate distribution, with optimal action existing in the separated narrow action intervals. Besides, changes in the opponent's strategy of the interaction will also change this dentate distribution. Such dynamic dentate action-reward distribution do not have an obvious action-incentive direction, and action ineria is prone to make the optimal solution non-optimal, which will fail to find an effective interaction pattern. Thus, it is not suggested to use Bellman equation to back-propagate the reward function as in static single-agent reinforcement learning in the scheduling problem. As illustrated in figure \ref{fig:fig3}, the effectiveness of the scheduling control for CAVs actually only depends on their arrival time at the crossing zone and their speed trajectories through the crossing zone, where vehicles have direct conflicts. There is actually more than one feasible speed trajectory in the preparation zone that can eventually produce the same result, i.e., the action control for CAVs is not strongly coupled in time dimension. Therefore, we adopt an immediate discriminator and a final discriminator to return short- and long-term interaction results, which can prevent curse of dimensionality result from joint space of environmental temporal evolution and inter-vehicle interactions.

For the convenient of presentation, base definitions are given below.
\begin{description}
    \item[Definition 1] \textbf{Predefined Trajectory.} To simplify the problem, this study focuses on the speed control of CAVs, and CAVs will drive along predefined trajectories through the intersection. These trajectories are expressed as $r_1, r_2, ..., r_j, ...$ which are $N_r$ in total.
    \item[Definition 2] \textbf{Conflict point.} When two predefined trajectories intersect in the crossing zone, their intersection point is called conflict point. Conflict points are expressed as $c_1, c_2, ..., c_i, ...c_{N_c}$. A set of conflict points passed by trajectory $r_j$ are denoted as $C_{r_j} := [c_{j1},c_{j2},...c_{ji},...]$. Correspondingly, the set of trajectories in conflict with $r_j$ are denoted as $R_{cj} = [r_{j1}, r_{j2}, ..., r_{ji},...]$.
    \item[Definition 3] \textbf{Reservation table.} The reservation table $M$ records the spatio-temporal occupancy of each CAV arriving at the crossing zone at the current speed, expressed in a table of $N_c\times N_t$, where $N_t$ indicates the furthest simulated step that will be recorded. When two CAVs request the same cell in the table simultaneously, it indicates a potential conflict between the two vehicles. It is noteworthy that the reservation table is only used to check the possible conflict of CAVs in the intersection area, and is not directly involved in intersection resource allocation.
    \item[Definition 4] \textbf{Active conflict \& Passive conflict.} When there is a conflict in the reservation table, the conflict is called a passive conflict for the party with priority and an active conflict for the other party.
\end{description}

\subsection{Reservation Mechanism}
\begin{figure}[h]
    \centering
    \includegraphics[width=0.6\textwidth]{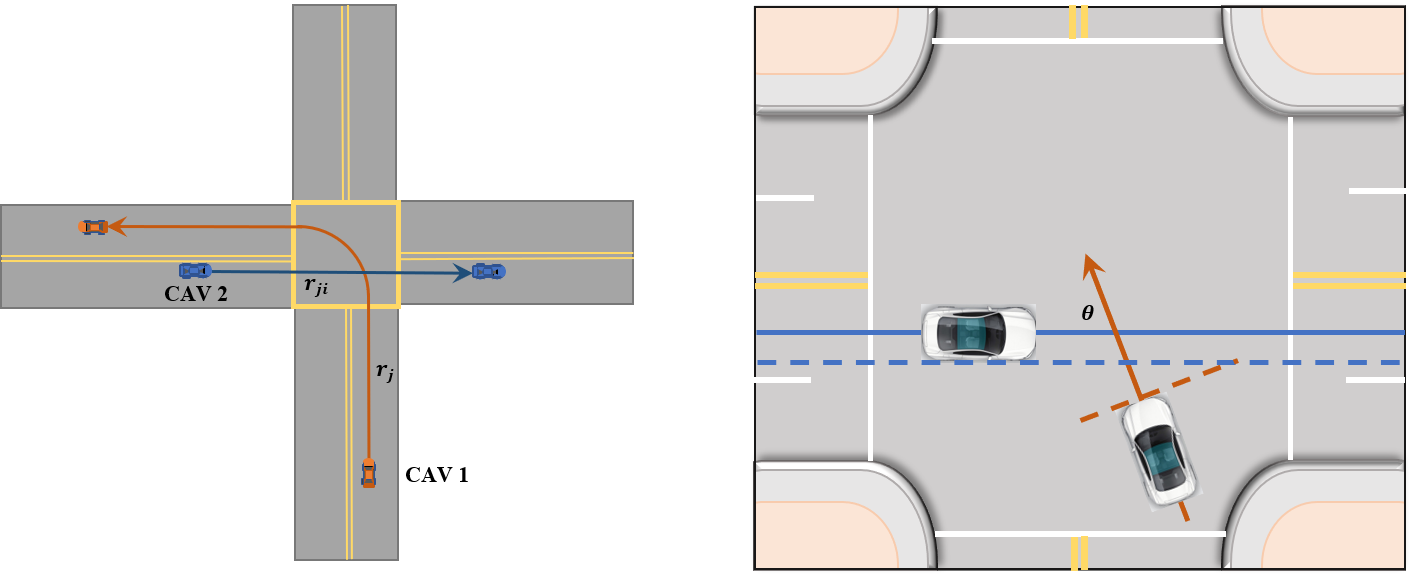}
    \caption{Explanation of the reservation mechanism}
    \label{fig:fig4}
\end{figure}

As shown in figure \ref{fig:fig4}, given a pre-defined trajectory $r_j$, it conflicts with $r_{ji} \in R_{cj}$ at $c_{ji} \in C_{r_j}$. The angle between the tangents of $r_j$ and $r_{ji}$ at $c_{ji}$ is defined as $\theta$. The CAV $H_{k1}$ along $r_j$ accesses $c_{ji}$ from start time $t_1$ to end time $t_2$, which is expressed as:

\begin{equation}
    t_1 = \frac{d_{ji}(H_{k1}) - w_{k1}/2 \cdot cot(\theta) - w_{k2}/2}{v_{k1}} \lesssim \frac{d_{ji}(H_{k1}) - \epsilon}{v_{k1}}
\end{equation}

\begin{equation}
    t_2 = \frac{d_{ji}(H_{k1}) + l_{k1} + w_{k1}/2 \cdot cot(\theta) + w_{k2}/2}{v_{k1}} \lesssim \frac{d_{ji}(H_{k1}) + l_{k1} + \epsilon}{v_{k1}}
\end{equation}

where $d_{ji}(H_{k1})$ is the distance from the front bumper of $H_{k1}$ to $c_{ji}$, $l_{k1}$, and $w_{k1}$ are the length and width of $H_{k1}$, respectively, $w_{k2}$ is the width of vehicle in conflict with $H_{k1}$. And $v_{k1}$ denotes the current speed of $H_{k1}$. To unify the expression and simplify the calculation, $\epsilon$ is introduced as a safe buffer.

The process of update reservation table $M$ is expressed as \ref{alg:alg1}

\begin{algorithm}[h]
    \caption{update $M$ and check for conflicts between CAVs at each step}
    \label{alg:alg1}
    \SetKwInOut{Input}{Inputs}
    \SetKwInOut{Output}{Outputs}
    \SetKwInOut{Definition}{Definition}
    \SetKwInOut{Initialize}{Initialize}
    \Definition{$H$: all vehicles currently in the intersection area that have not yet passed.\\
                $H'$: all vehicles in conflicts}
    \Input{$H$}
    \Output{$M$,$H'$}
    \Initialize{Clear all cells in $M$; $H' = \{\}$}
    \For{$H_{i}$ in $H$}{
        $r_j \leftarrow$ the trajectory $H_{i}$ drives along\\
        $R_{cj} \leftarrow$ the trajectories in conflict with $r_j$\\
        \For{$r_{jk}$ in $R_{cj}$}{
            $d_{jk}(H_i) \leftarrow$ the distance from the front bumper of $H_i$ to $c_{jk}$\\
            $v_{H_i} \leftarrow$ the current speed of $H_i$ given by simulator\\
            $t1 \leftarrow$ Equation (1)\\
            $t2 \leftarrow$ Equation (2)\\
            $t1 = max(min(t1,T_r),0)$\\
            $t2 = max(min(t2,T_r),0)$\\
            $idx \leftarrow$ the index of $c_{jk}$ in $M$\\
            \For{cell in $M(idx,t1:t2)$}{
                \If{len(cell) > 0}{
                    $H'$.add($H_i$)\\
                    $H_c \leftarrow$ the vehicles in conflict with $H_i$ in cell\\
                    $H'$.union($H_c$)\\
                }
                cell.add($H_i$)\\
            }            
        }
    }
    \textbf{Return} $M$, $H'$
\end{algorithm}

\subsection{Hierarchical Adversarial Learning}
\begin{figure}[h]
    \centering
    \includegraphics[width=0.8\textwidth]{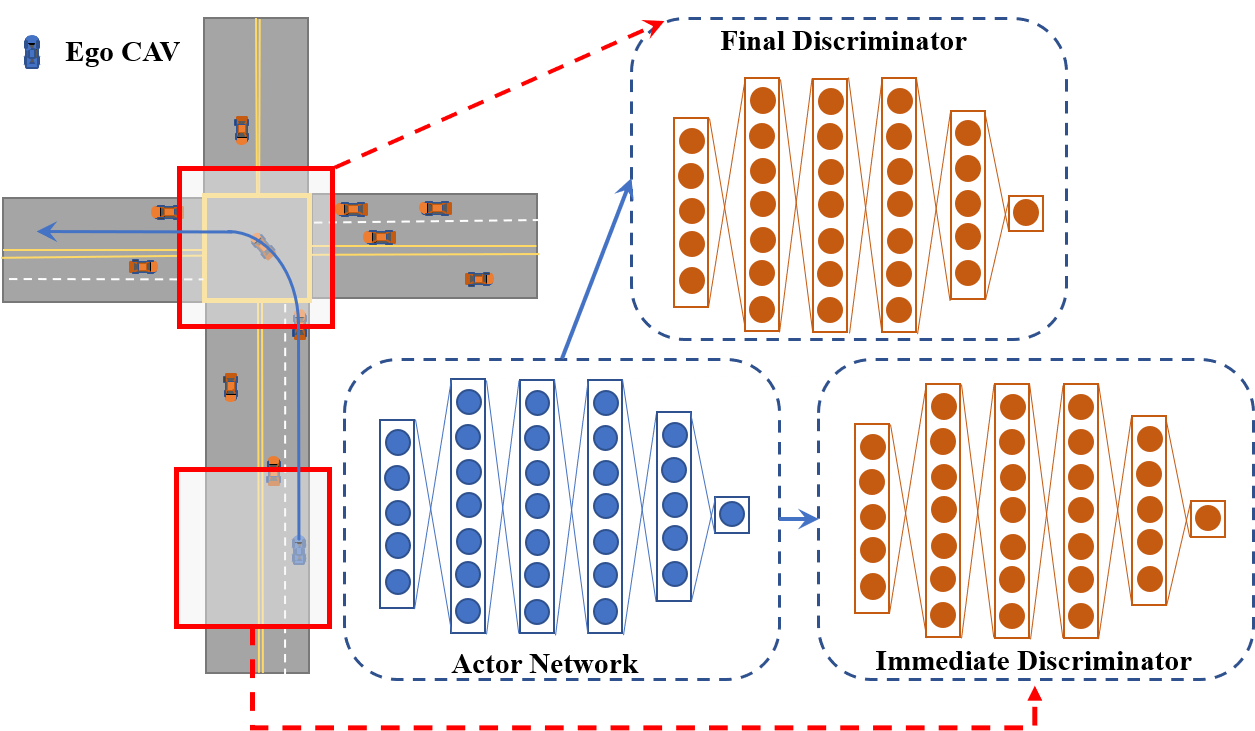}
    \caption{The general framework of hierarchical adversarial learning}
    \label{fig:fig5}
\end{figure}

The hierarchical adversarial learning includes an actor network, an immediate interaction discriminator, and a final trajectory discriminator as shown in figure \ref{fig:fig5}. The actor network produces actions for CAVs in the preparation zone at each decision step. The immediate interaction discriminator check the immediate conflicts to break symmetric decision dilemma between pairs of homogeneous agents and teach them the correct right-of-way at each action step. The trajectory discriminator judges whether there will be conflicts eventually and feedbacks the long-term reward to encourage farsighted behavior.

The inputs to the actor network include global observation of environment $G_t$ and motion state of ego vehicles $x_t(H_i)$: $s_t(H_i) = [x_t(H_i), G_t]$. The global observation $G_t$ records the distribution of vehicles on $N_r$ pre-defined trajectories: $G = [g_{r_1}, g_{r_2}, ..., g_{r_{N_r}}]$, where $g_{r_j}$ divides part of $r_j$ from entering the preparation zone to leaving the crossing zone into segments with a certain length, and fills the vehicle's movement attributes into the corresponding cell based on the position of each vehicle's front bumper, including distance to the stop line (negative after crossing the stop line), speed, angle, and whether it conflicts with ego CAV ($1$ for conflict, $0$ for no conflict), and all these information is aggregated into $G_t$. The ego state consists of three one-hot vectors: $x_t(H_i) = [x_r(H_i),x_d(H_i),x_v(H_i)]$, which indicate the trajectory $H_i$ drives on, the value and interval of the distance to the exit point, and the value and the interval of the current speed, respectively. All elements in the state vector $s_t(H_i)$ are normalized to the interval from $-1$ to $1$. The environment state $G_t$ is extracted by a three-layer MLP then concatenate with ego state $x_t(H_i)$, and produces output after the other three-layer MLP. After the activation layer, the output is projected to $[a_{min},a_{max}]$, where $a_{min}, a_{max}$ are the maximum deceleration and acceleration of the ego CAV.

The state-action pair ($s_t, a_t$) is fed to the immediate interaction discriminator and final trajectory discriminator, respectively. The immediate interaction discriminator involves a conflict discriminator, an active conflict discriminator, and a passive conflict discriminator. when the action $a_t$ leads to a conflict in the reservation table at next step, the true label of the conflict discriminator $L_C$ equals to $1$, otherwise $0$. If conflict exists, the priority is determined by three rules:
\begin{description}
    \item[Rule 1] When two vehicles $H_i, H_j$ conflict, check the last-step reservation table. If at the previous moment, the step $H_j$ expected to arrive at the conflict point is later than the step $H_i$ expected to leave the conflict point, then $H_i$ has priority, i.e., unreasonable acceleration $H_j$ or deceleration $H_i$ cause new conflict, the vehicle that was originally first has priority.
   
    If Rule 1 is not satisfied (i.e., there is also a conflict between $H_i$ and $H_j$ at the previous step), the priority is given by Rule 2 and Rule 3.
    \item[Rule 2] When $H_i$ and $H_j$ are in adjacent edges, the vehicle on the right has priority.
    \item[Rule 3] When $H_i$ and $H_j$ are in opposite edges, the vehicle going straight has priority.
\end{description}

After priority is determined, the party with priority in the conflict is called to have a passive conflict, and the other party is called to have an active conflict. When the active conflict occurs, the true label of active conflict discriminator $L_{AC}$ equals to $1$, otherwise $0$. And the same holds true for the true label of passive conflict discriminator $L_{PC}$. And the corresponding predicted labels of the three discriminators are denoted as $\hat{L}_C,\hat{L}_{AC}',\hat{L}_{PC}'$. According to the definitions, $L_{C}, L_{AC}, L_{PC}$ should satisfy: $L_{C} = L_{AC}+L_{PC}$, thus the final predicted labels of active conflict and passive conflict are expressed as:
\begin{equation}
    \hat{L}_{AC} = \alpha \hat{L}_{AC}' + (1-\alpha) (\hat{L}_{C} - \hat{L}_{PC}) 
\end{equation}

\begin{equation}
    \hat{L}_{PC} = \alpha \hat{L}_{PC}' + (1-\alpha) (\hat{L}_{C} - \hat{L}_{AC}) 
\end{equation}
where $\alpha$ is the hyper-parameter that weighs the labels from different discriminators, and the values of $\hat{L}_{AC},\hat{L}_{PC}$ are truncated within $[0,1]$. Apart from immediate conflicts, the final trajectory discriminator predicts the final conflicts as the ego CAV passes through the crossing zone along its action trajectory. It informs actor network the potential result from each-step action and helps it form a far-sighted decision-making pattern. The state-action trajectory is given as: $s_0 \rightarrow a_1 \rightarrow s_1 \rightarrow a_2 \rightarrow ... \rightarrow a_{m-1} \rightarrow s_m \rightarrow a_{m} \rightarrow \delta_c$, where $s_0$ and $s_{m}$ is the first and last state of ego CAV in the preparation zone, and $a_t$ is the action at the corresponding moment, then $\delta_c$ records whether ego vehicle along this trajectory collides with others in the crossing zone. A memory palace $D$ is adopted to store experiences from all CAVs and train these discriminators. Take the aforementioned trajectory as an example, the update of memory palace will be:
\begin{equation}
    D = D \cup \{(s_0, a_0, L_{AC,0}, L_{PC,0},\delta_c), (s_1, a_1, L_{AC,1}, L_{PC,1},\delta_c),...,(s_m, a_m, L_{AC,m}, L_{PC,m},\delta_c)\}
\end{equation}

Given an action the actor network generates for $H_i$ at $t$ step, $a_t \in [a_{min}, a_{max}]$, the loss of actor network is expressed as:
\begin{equation}
    l = \beta (1-\hat{L}_{PC})(1-\hat{L}_{AC})(a_{max} - a_t) + (1-\beta)\hat{\delta}_C\big[(a_{max}-a_t)\hat{L}_{PC} + (a_t-a_{min})\hat{L}_{AC}\big]
\end{equation}

\begin{equation}
    \beta = \gamma \frac{d_t - d_{min}}{d_{max} - d_{min}}
\end{equation}
The first term of Equation 6 is an efficiency term that encourages the CAV to accelerate in the conflict-free situation, while the second term is a safety term that encourages the CAV to achieve conflict avoidance by proper acceleration or deceleration. The weight $\beta$ is a potential energy term -- the closer the CAV is to the conflict points, the more important safety term is, and $d_t$ is the CAV's current distance from the stop line, $d_{min}$ and $d_{max}$ are the maximum and minimum control boundaries. The term $\gamma$ is a discount factor, which is set near to $1$ in the early stage of training to improve efficiency, and decreases as the training process to focus more on conflict avoidance. 

When two CAVs conflict in the reservation table, the party with priority has $\hat{L}_{PC}\rightarrow 1$ and $\hat{L}_{AC}\rightarrow 0$, then the loss function will induces $a_t \rightarrow a_{max}$ to make the CAV go first, while the other party are intend to decelerate by $a_t \rightarrow a_{min}$. When a CAV has both active conflict with front vehicles and passive conflict with rear vehicles, it has $\hat{L}_{PC}\rightarrow 1$ and $\hat{L}_{AC}\rightarrow 1$ and receives a nearly constant loss $(1-\beta)\hat{\delta}_C(a_{max}-a_{min})$ from the safety term. It will be motivated by the efficiency term to accelerate appropriately and wait for the movements of front and rear vehicles to deconflict. The prediction label of trajectory discriminator $\hat{\delta}_c$ equals to $1$ when collisions occur finally and $0$ otherwise, it encourages the multiple parties in a conflict ring to adjust their speed differently to prevent the deadlocks and final collisions.

To further regulate the behavior of CAVs, state maintenance and action mask mechanisms are applied in the decision process. The state maintenance requires CAVs to maintain the driving speed of the previous moment when they have no conflict with others and their speed is above a certain threshold. The mechanism helps to reduce the complexity of interaction. And the action mask prevents CAVs from adopting actions that would cause new conflicts in the reservation table and forces them to maintain the velocity state of the previous moment, and this mechanism allows the model to focus on resolving existing conflicts since entering the intersection area and find a efficient scheduling. In the context of pure CAVs using homogeneous agent control, the ego CAV can effectively predict the actions of other vehicles within a finite number of steps and update its own stored reservation table. In this process, the action mask mechanism checks future conflicts for several steps and allows for temporary conflicts resulting from swaps in the order of passage among CAVs.

\section{Experiments}
\subsection{Experiment Configuration}
In this study, the proposed model and baselines are evaluated on a widely-used microscopic traffic simulator \textit{Simulation of Urban Mobility} (SUMO). The experiments are conducted using Python 3.8 on a computer with a 12-core i7-12700KF @4.9GHz/RAM: 32GB processor and NVDIA 3080. The simulation scenario is a four-way-six-lane intersection containing a right-turn lane, a straight lane, and a left-turn lane in each direction. All CAVs are spawned at $200$-m from the stop lines and are controlled after they enter the preparation zone. Each epoch of simulation lasts for $300$ seconds in the training process and $10000$ seconds in the test process. The parameters of the model and simulated environment are listed in Table \ref{tab:tab1}. 

In order to adapt the model to various traffic pressure, we cycle through the training process with different intensities of traffic flow ranging from $6000$veh/h to $9600$veh/h, and each cycle includes 15 epochs, where the flow intensity of the first 9 epochs is $6000-7200$veh/h, and that of epoch 9 to 12 and the last 3 epochs are $7200-8400$veh/h, $8400-9600$veh/h, respectively. The flow rate in the test is divided into low, medium and high traffic conditions, respectively $5400$veh/h, $7200$veh/h, $9000$veh/h. To better evaluate how the model behaves in reality, the traffic flows are loaded unevenly in all directions and changes over time, with the overall vehicles arrivals obeying as Poisson distribution.

\begin{table}[htb]
    \centering
    \begin{tabular}{lll}
    \toprule[1.5pt]
    \makebox[0.15\textwidth][l]{Categories}    & \makebox[0.4\textwidth][l]{Parameters}  & \makebox[0.3\textwidth][l]{Values} \\ 
    \midrule[1.0pt]
    Environment Parameters  &  Simulation duration (training)  &   300s/epoch\\
                            &  Simulation duration (testing)   &   10000s/epoch\\
                            &  Control distance                &   100m\\
                            &  Simulation step                 &   0.1s\\
                            &  Action step of CAVs             &   0.2s\\
                            &  Length of CAVs                  &   5.0m\\
                            &  Width of CAVs                   &   1.8m\\
                            &  Maximum speed of CAvs           &   15m/s\\
                            &  Maximum acceleration of CAvs    &   4m/s2\\\midrule[1.0pt]
    Model Parameters        &  Maximum number of training epochs   & 100\\
                            &  Batch size                          & 256\\
                            &  Buffer size                         & $10^6$\\
                            &  Actor learning rate                 & $10^{-5}$\\
                            &  Discriminator learning rate         & $10^{-5}$\\
                            &  Optimizer                           & Adam\\
                            &  Weight alpha in immediate discriminator & 0.6\\
                            &  Initial weight gamma                    & 0.2\\
                            &  Maximum recorded period of reservation table   & 100s\\
    \bottomrule[1.5pt]
    \end{tabular}
    \caption{Parameter Configuration}
    \label{tab:tab1}
\end{table}

\subsection{Metrics}

To systematically evaluate the performance of the proposed model and benchmarks in terms of safety, efficiency, fairness, and energy saving, the following five metrics are adopted in the experiments:

\begin{itemize}
    \item[(1)] \textbf{Passing Ratio}. Number of CAVs passing through the intersection divided by the total number of CAVs.
    \item[(2)] \textbf{Stopping Ratio}. Number of CAVs stop before the stop lines divided by the total number of CAVs.
    \item[(3)] \textbf{Average Travel Time}. The average travel time of CAvs from entering the preparation zone to leaving the crossing zone.
    \item[(4)] \textbf{Fairness}. The standard deviation of CAVs' travel time is applied to measured the fairness among different CAVs.
    \item[(5)] \textbf{Fuel Consumption}. The average fuel consumption of CAVs. The indicator reflects the degree and frequency of acceleration and deceleration of CAVs. 
\end{itemize}

\subsection{Baselines}
Seven methods are used as comparison with the proposed model.
\begin{itemize}
    \item[(1)] \textbf{Fixed-Timing Signal Control (FT)}. A traditional signal control method. The cycle length and the phase duration do not change during each epoch. For low, medium, and high traffic test scenarios, the signal cycles are set to $60$s, $90$s, and $120$s, respectively.
    \item[(2)] \textbf{Longest Queue First Siganl Control (LQF)}. A pressure-based adaptive signal control technique. LQF gives the green light to the phase with the longest queue.
    \item[(3)] \textbf{First-Come-First-Serve (FCFS)}. The CAV that arrives first in the crossing zone is the first to cross the intersection, and when a CAV is not in conflict with the CAVs with higher priority in the crossing sequence, it is allowed to pass through the intersection simultaneously.
    \item[(4)] \textbf{First-Come-First-Serve with Platooning (Platoon)}. When the gap between the CAV and preceding CAV is shorter than a certain threshold and the size of platoon where the preceding CAV is located (8 in this study), the CAV joins the platoon. The first CAV of the platoon is called the leader CAV, and the platoon cross the intersection according to the priority of the leader CAV.
    \item[(5)] \textbf{Twin Delayed Deep Deterministic Policy Gradient (TD3)}. One of the advanced reinforcement learning model using actor-critic framework. Double Q Network is applied to prevent overestimation.
    \item[(6)] \textbf{Soft Actor-Critic (SAC)}. One of the advanced reinforcement learning using actor-critic framework. The entropy term encourages agent to explore more actions in the early stage, and fuller exploration enables SAC agent to find better solutions in complicated situations. Its reward curves rise slowly but steadily, and can usually reach a high value.
    \item[(7)] \textbf{DCL-AIM}. One of the advanced reinforcement learning framework for AIM issue, independent and coordination strategy was adopted to control CAVs in DCL-AIM. The granularity of the rasterization is set to $12$.
\end{itemize}

\subsection{Results and Analysis}
\subsubsection{Model Convergence}

\begin{figure}[htb]
    \centering
    \includegraphics[width=\textwidth]{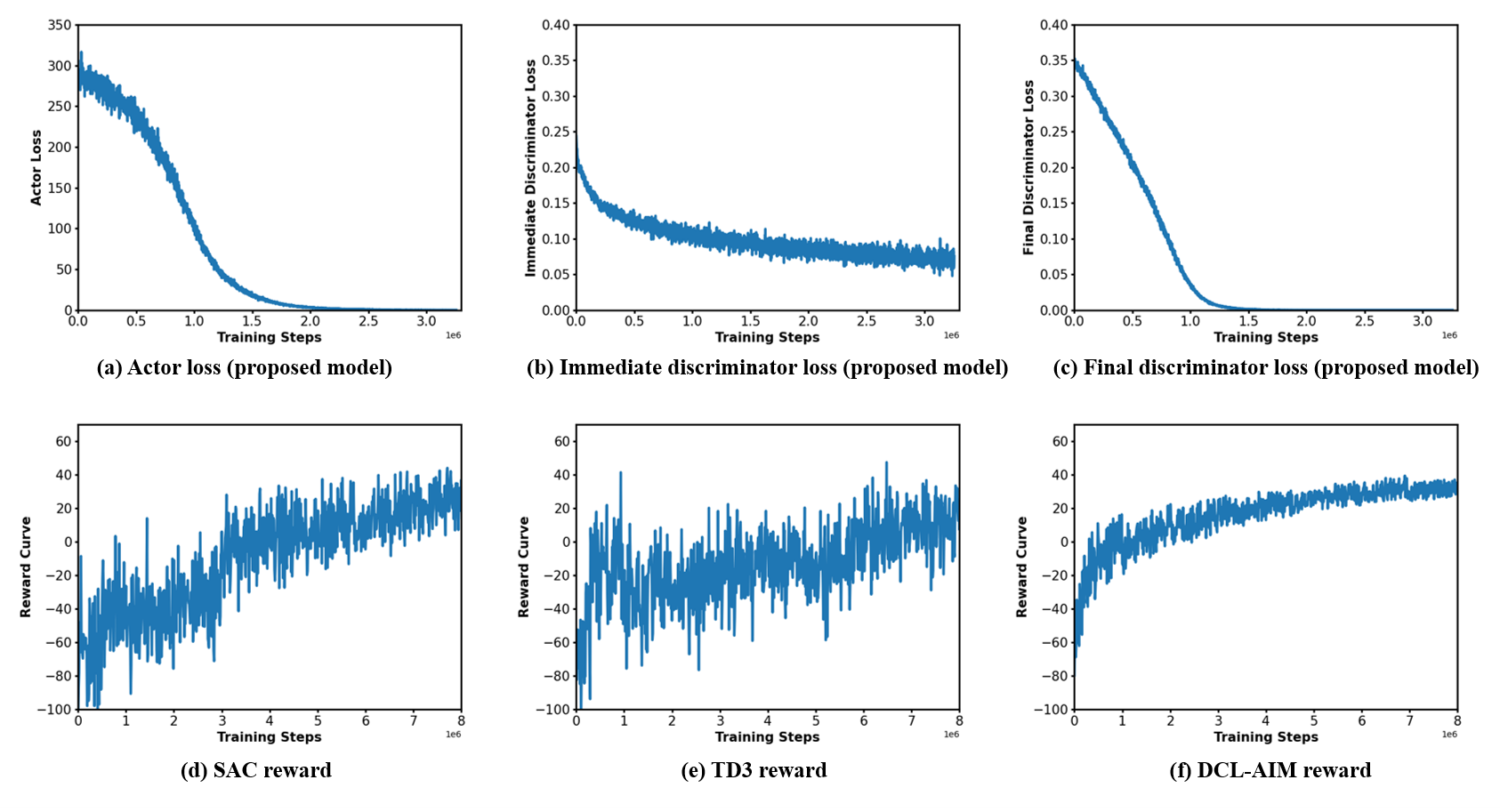}
    \caption{The convergence of proposed model and learning-based baselines}
    \label{fig:fig6}
\end{figure}

The first row of diagram in Figure \ref{fig:fig6} exhibits the loss curves of the actor network, the immediate discriminator, and the final discriminator in the proposed model. The entire training process includes 3.3 million steps, and the proposed model achieves nice convergence, the loss of both actor network and the final discriminator converge to zero after 2.0 and 1.5 million steps, respectively, and the loss of immediate discriminator reaches stable at about 0.07. According to the Equation (6), the zero value of actor loss means that the CAV can accelerate to avoid being clashed potentially in a passive conflict and slow down to yield to the opponent CAV in an active conflict. When a CAV is both in a passive conflict and an active conflict in some cases, $[(a_{max}-a_t)\hat{L}_{PC} + (a_t -a_{min})\hat{L}_{AC}]$ in Equation (6) will be constant, and the zero value of actor loss means the proposed model enables the CAV to eventually achieve a complete collision free in the crossing zone, which makes $\hat{\delta}_C$ zero. 

The reward design for the three reinforcement learning model (i.e., SAC, TD3, DCL-AIM) is the same. A bonus of $100$ is awarded when the CAV can safely pass through the intersection without collision, and $-100$ when a collision occurs while crossing the crossing zone. To encourage CAVs to pass through the intersection efficiently, a weak penalty is awarded for each step before their passage, with a penalty of $-0.2$ when the CAV does not have potential conflict with other opponents in the reservation table, and $-1$ otherwise, which is to guide CAVs to prepare in advance to avoid collisions. The training process of these RL model is longer than the proposed model, which has 8 million steps. The rewards of the SAC model are significantly improved in the first 5 million steps, and culminate in an average reward of about 25 per vehicle. The reward curve for TD3 rises faster in the early stage, but the final convergence value is slightly lower than that of SAC as SAC has richer exploration of the interaction pattern. The reward curves of both SAC and TD3 have relatively strong oscillations in the rise, which is due to the fact that it is difficult to completely avoid conflicts between ego CAV and the opponent CAVs in homogeneous-agent interaction, and occasional accidents can bring large negative rewards and affect the stability of learning. By introducing the coordination state and collision-free mechanism, DCL-AIM shows a clearer and more stable upward trend, the diagram in figure \ref{fig:fig6} gives the average reward obtained per CAV under the combined effect of both individual and coordinated decisions.

It can be seen from figure \ref{fig:fig6} that the convergence of the proposed model outperforms the three selected comparison reinforcement learning, which validates the point made in the previous chapter that the hierarchical adversarial learning allows for a more concise optimization space of CAV's trajectory and can acquire better interaction results.

\subsubsection{Model Comparison}

\begin{table}[h]
    \caption{Results of test scenarios at low traffic volumes}
    \begin{tabular}{lccccc}
    \toprule[1.5pt]
                      & \makebox[0.12\textwidth][c]{PR}  & \makebox[0.12\textwidth][c]{SR}     & \makebox[0.12\textwidth][c]{ATT}  & \makebox[0.12\textwidth][c]{DTT} & \makebox[0.12\textwidth][c]{AFC}     \\ 
                      \midrule[1.0pt]
    FT  (signalized)    & $100.0\pm 0.0$   & $92.4\pm 6.3$   & $31.12 \pm 9.53$  & $24.69 \pm 3.52$   & $39.61 \pm 7.65$\\
    LQF (signalized)    & $100.0\pm 0.0$   & $97.3\pm 2.2$   & $13.25 \pm 0.72$  & $16.76 \pm 3.60$   & $21.23 \pm 4.08$\\
    FCFS                & $100.0\pm 0.0$   & $84.7\pm 6.5$   & $46.63 \pm 8.17$  & $43.12 \pm 2.05$   & $43.35 \pm 6.15$\\
    FCFS (platoon)      & $100.0\pm 0.0$   & $79.5\pm 4.8$   & $27.68 \pm 3.01$  & $26.70 \pm 4.11$   & $32.74 \pm 6.21$\\
    TD3                 & $88.6 \pm 4.3$   & $2.3 \pm 0.6$   & $8.26  \pm 0.25$   & $3.68 \pm 0.66$    & $65.37 \pm 10.39$\\
    SAC                 & $89.1 \pm 5.0$   & $2.4 \pm 0.9$   & $8.08  \pm 0.24$  & $3.51 \pm 0.92$    & $59.17 \pm 9.23$\\
    DCL-AIM             & $100.0\pm 0.0$   & $4.6 \pm 1.3$   & $12.51 \pm 0.63$  & $5.92 \pm 1.74$    & $49.60 \pm 7.48$\\
    D-SAL(proposed)     & $100.0\pm 0.0$   & $2.7 \pm 0.4$   & $9.13  \pm 0.27$  & $3.24 \pm 0.51$    & $46.79 \pm 5.23$\\
    \bottomrule[1.5pt]
    \end{tabular}
    \label{tab:tab2}
    \begin{tablenotes}
        \item[*] (\textbf{PR}: Passing Ratio(\%); \textbf{SR}: Stops Ratio(\%); \textbf{ATT}: Average Travel Time (s); \\ \textbf{DTT}: Standard Deviation of Travel Time(s); \textbf{AFC}: Average Fuel Consumption(mL/veh))
    \end{tablenotes}
\end{table}

\begin{table}[h]
    \caption{Results of test scenarios at medium traffic volumes}
    \begin{tabular}{lccccc}
    \toprule[1.5pt]
                      & \makebox[0.12\textwidth][c]{PR}  & \makebox[0.12\textwidth][c]{SR}     & \makebox[0.12\textwidth][c]{ATT}  & \makebox[0.12\textwidth][c]{DTT} & \makebox[0.12\textwidth][c]{AFC}     \\ 
                      \midrule[1.0pt]
    FT  (signalized)    & $100.0\pm 0.0$   & $98.8 \pm 0.7$   & $47.03 \pm 7.99$   & $56.83\pm5.22$   & $42.37 \pm 4.16$\\
    LQF (signalized)    & $100.0\pm 0.0$   & $99.2 \pm 0.6$   & $14.61\pm 1.14$    & $18.72\pm2.41$   & $29.11 \pm 3.76$\\
    FCFS                & $100.0\pm 0.0$   & $93.5 \pm 2.8$   & $53.22 \pm 10.57$  & $36.88\pm4.27$   & $62.23\pm 5.21$\\
    FCFS (platoon)      & $100.0\pm 0.0$   & $89.3 \pm 3.1$   & $37.60 \pm 7.24$   & $33.62 \pm2.95$  & $59.12 \pm 5.13$\\
    TD3                 & $81.5 \pm 2.6$   & $3.4 \pm 0.9$    & $8.30 \pm 0.46$    & $3.92 \pm 0.58$  & $62.56\pm7.82$\\
    SAC                 & $82.0 \pm 2.5$   & $3.0 \pm 0.8$    & $8.11 \pm 0.38$    & $3.65 \pm 0.78$  & $57.19\pm8.74$\\
    DCL-AIM             & $100.0\pm 0.0$   & $5.3 \pm 1.1$    & $13.59 \pm 0.76$   & $5.88 \pm 2.03$  & $52.63 \pm 6.25$\\
    D-SAL(proposed)     & $100.0\pm 0.0$   & $4.5 \pm 0.6$    & $9.31\pm0.69$      & $3.31 \pm 0.56$  & $45.46 \pm 5.18$\\
    \bottomrule[1.5pt]
    \end{tabular}
    \label{tab:tab3}
\end{table}

\begin{table}[h]
    \caption{Results of test scenarios at high traffic volumes}
    \begin{tabular}{lccccc}
    \toprule[1.5pt]
                      & \makebox[0.12\textwidth][c]{PR}  & \makebox[0.12\textwidth][c]{SR}     & \makebox[0.12\textwidth][c]{ATT}  & \makebox[0.12\textwidth][c]{DTT} & \makebox[0.12\textwidth][c]{AFC}     \\ 
                      \midrule[1.0pt]
    FT  (signalized)    & $100.0\pm 0.0$   & $99.8\pm0.2$   & $54.03 \pm 2.19$   & $66.29 \pm 5.75$   & $47.93\pm3.21$\\
    LQF (signalized)    & $100.0\pm 0.0$   & $100.0\pm0.0$  & $24.26 \pm 2.31$   & $24.26 \pm 1.59$   & $40.92\pm3.88$\\
    FCFS                & $100.0\pm 0.0$   & $98.2\pm1.6$   & $69.84 \pm 7.02$   & $74.81 \pm 6.06$   & $55.03\pm5.60$\\
    FCFS (platoon)      & $100.0\pm 0.0$   & $99.2\pm0.8$   & $63.29 \pm 8.76$   & $43.50 \pm 5.13$   & $52.13\pm4.47$\\
    TD3                 & $73.9 \pm 6.4$   & $3.7\pm1.4$    & $8.40 \pm 0.33$    & $4.03 \pm 0.85$    & $61.65\pm5.83$\\
    SAC                 & $72.4 \pm 5.3$   & $2.9 \pm1.1$   & $8.08 \pm 0.75$    & $3.61 \pm 0.97$   & $56.32\pm6.65$\\
    DCL-AIM             & $100.0\pm 0.0$   & $22.6\pm3.5$   & $26.95 \pm 3.20$   & $7.02 \pm 1.92$   & $54.83\pm5.24$\\
    D-SAL(proposed)     & $100.0\pm 0.0$   & $16.5\pm3.7$   & $15.28 \pm 1.17$   & $3.60 \pm 0.48$   & $45.43 \pm4.32$\\
    \bottomrule[1.5pt]
    \end{tabular}
    \label{tab:tab4}
\end{table}

\begin{figure}[htb]
    \centering
    \includegraphics[width=\textwidth]{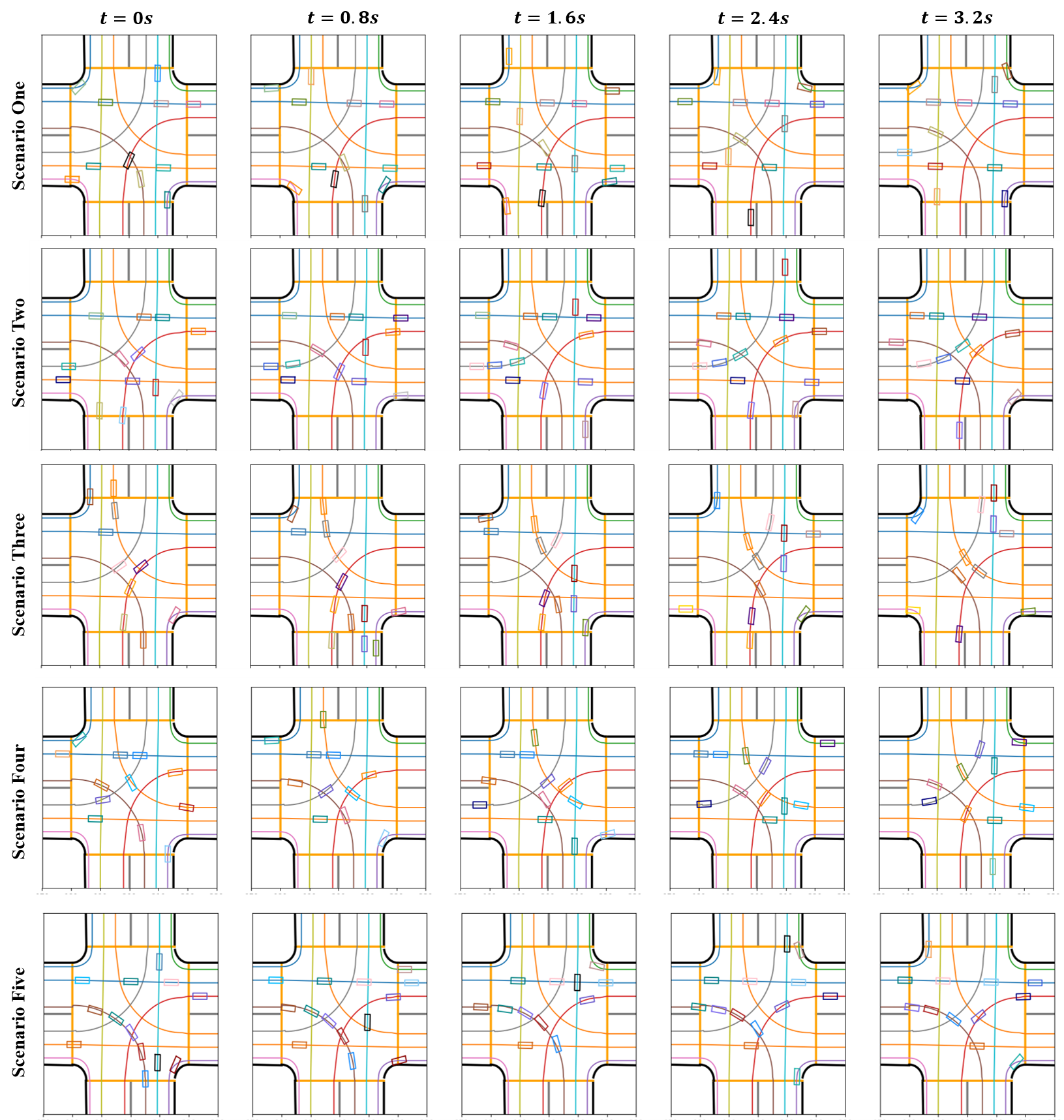}
    \caption{Several Interaction Cases}
    \label{fig:fig7}
\end{figure}

As presented in Table \ref{tab:tab2}-\ref{tab:tab4}, the SAC and TD3 model achieve the shortest pass time and the least stop ratio for all three traffic conditions. However, when applying such general reinforcement learning frameworks to AIM problem, the feedback of vehicular collisions as rewards are not guaranteed to produce effective constraints on the agent's behavior. In highly complex intersection interaction scenarios, the agent's policy may lose stability and eventually not converge due to frequent changes in the adversary agents' policy, or take a long time to converge with a reasonable training policy and reward mechanism. In the experiments, the TD3 and SAC can only achieve a pass rate of $88.6\%/81.5\%/73.9\%$, $89.1\%/82.0\%/72.4\%$, respectively. The proposed model achieves excellent collision avoidance by simplifying the interaction of agents in the time dimension while satisfying the short-term conflict-free and long-term collision-free conditions. Under the premise of safety first, the model realise the most efficient intersection passage and the least stopping ratio. It has close performance to the two aforementioned generic RLs and outperforms the RL framework designed for AIM, DCL-AIM. Generally, the improvement of the proposed model over the existing state-of-the-art method DCL-AIM is $27.0\%$, $31.5\%$,$43.3\%$ under low, medium, high traffic pressures, respectively.

In terms of fairness, First-Come-First-Serve seems to have more equitable principle, but it does not perform as well as it could. The method does not make full use of intersection capacity when some conflicting CAVs have adjacent priorities, and some CAVs that can cross simultaneously in the non-conflicting direction are forced to wait before the stop line, leading to large variations in efficiency at different time and with different short-time flows, thus enlarge the travel time differences for CAVs arriving at different instants. The learning-based approach, on the other hand, is incentivized by efficiency terms and reduces forced stops and waits, resulting in more equity in terms of differences in travel time. With regard to fuel consumption, it is mainly caused by two reasons, the stop-and-go behavior of CAVs before crossing the intersection in the signal-based and priority-based management, and the acceleration and deceleraion in adjustment of arrival time and crossing speed in the action-based method. Without the inclusion of fuel consumption feedback, the fuel consumption of CAV in the action-based method is more than that in the signal-based method. Among these action-based method, the proposed mode has the least consumption due to less speed adjustment in the entire trajectory. In addition, the gap in fuel consumption between the action-based and signal-based methods is shrunk as traffic volume rise.

\subsubsection{Interaction Pattern}

We find that CAVs tend to pass the crossing zones at unsignalized intersection with some specified patterns, some of which are enumerated in figure \ref{fig:fig7}. Each scenarios in figure \ref{fig:fig7} contains snapshots with a total interval of $3.2$s. In Scenario one, CAVs in the north-south, south-north, east-west, west-east interleave through the intersection, and CAVs in other directions intersperse through the gaps between CAVs in those directions. The conflict points between left-turn lanes are more concentrated than those between straight lanes, which are more likely to generate conflicts and deadlocks in unsignalized control. When crossing these conflict points, CAVs tend to adopt strategies like rhythmic control. Each rhythm cycle generally contains three or four CAVs turning left in sequence, as shown in Scenario two to Scenario four in \ref{fig:fig7}. The crossing order among them is usually organized as: CAVs in two non-conflicting directions turn left simultaneously, and then CAVs in the two other non-conflicting directions turn left simultaneously, or four CAVs turn left in a clockwise or counterclockwise direction in sequence. The combined patterns of straight and left turns are not significant at low traffic volumes, but arise when traffic volumes are elevated, i.e., various specified rhythmic patterns of movements arise between CAVs in all lanes, instead of separate crossing patterns between straight and left-turn CAVs respectively. In high traffic flows, platoon formation occurs spontaneously among CAVs in the same direction just like platoon-based method and LQF control, but unlike these two method, the spontaneous formation will enlarge the headway with front CAV to enable the upcoming CAVs in the conflicting direction to pass as shown in Scenario Five\ref{fig:fig7}, reducing the waiting time and unnecessary slowdown and stop of conflicting CAVs, and further improving the efficiency of passage.

\section{Conclusion}

In this study, we develop a non-RL learning-based framework for addressing AIM problem with pure CAVs. The framework applies hierarchical adversarial learning to regulate the instant and farsighted behaviors of CAVs with immediate discriminators and a final discriminator. To the best of our knowledge, this study is one of the first attempts to decouple the temporal dimension from learning-based methods, and enable CAVs to concentrate more on the immediate interaction and the final results of the interaction, which is found to be robust to conflict avoidance. To balance safety and efficiency, a dynamic potential energy term is used to weigh the importance of efficiency at different distances from the crossing zone and on different stages of training process. Besides, two important mechanisms -- state maintenance and action mask enable the actor network to find the optimal policy more stably in the interaction.

The experiments to evaluate and validate the proposed model are configured at a four-way-six-lane intersections, the results manifest that the proposed model: (1) has satisfactory convergence on the losses of actor network and discriminator networks. (2) can navigate CAVs to intersperse and interact efficiently during intersection passage. Besides, several methods commonly used for intersection scheduling management are compared with the proposed model, the comparisons show that the proposed model can achieve the fastest passage while ensuring safety.

However, since the action-based method involves frequent acceleration and deceleration, fuel consumption and comfort may inferior to some priority-based and signal-based methods, which need to be further enhanced in the future research, possibly by involving appropriate relevant objectives in the training.

\bibliographystyle{unsrt}  
\bibliography{references}

\end{document}